\documentclass{aa}
\usepackage{graphics}

\def\la{\lower.5ex\hbox{$\; \buildrel < \over \sim \;$}}
\def\ga{\lower.5ex\hbox{$\; \buildrel > \over \sim \;$}}

\begin{document}      
   
   \title{A holistic view on ram pressure stripping in the Virgo cluster}
   \subtitle{The first complete model-based time sequence}

   \author{B.~Vollmer}

   \offprints{B.~Vollmer: bvollmer@astro.u-strasbg.fr}

   \institute{CDS, Observatoire astronomique de Strasbourg, 11, rue de l'universit\'e,
	      67000 Strasbourg, France}

   \date{Received / Accepted}

   \authorrunning{B.~Vollmer}
   \titlerunning{A holistic view on ram pressure stripping in the Virgo cluster}

\abstract{Based on a comparison of dynamical models with observations of the interstellar gas in 6  Virgo cluster spiral
galaxies a first complete ram pressure stripping time sequence has been established. The observational characteristics 
of the different stages of ram pressure stripping are presented. The dynamical models yield the 3D velocity vectors of 
the galaxies,  peak ram pressures, and times to peak ram pressure. In the case of a smooth, static, and spherical 
intracluster medium, peak ram pressure occurs during the galaxy's closest approach to the cluster center, i.e. when the 
galaxy's velocity vector is perpendicular to its distance vector from the cluster center (M87). Assuming  this condition 
the galaxy's present line-of-sight distance   and its 3D position during peak ram pressure can be calculated. The linear 
orbital segments derived in this way together with the intracluster medium density distribution derived from X-ray 
observations give estimates of the ram pressure that are on average  a factor of 2 higher than derived from the dynamical 
simulations for NGC~4501, NGC~4330, and NGC~4569. Resolving this discrepancy would require either a 2 times higher 
intracluster medium density than derived from X-ray observations, or a 2 times higher stripping efficiency than assumed 
by the dynamical models. Compared to NGC~4501, NGC~4330, and NGC~4569, NGC~4388 requires a still 2 times higher local 
intracluster medium  density or a direction which is moderately different from that  derived from the dynamical model. 
A possible scenario for the dynamical evolution of NGC~4438 and M~86 within the Virgo cluster is presented.
}

\maketitle

\section{Introduction \label{sec:introduction}}

The evolution of a spiral galaxy is strongly affected by its environment giving rise to the density-morphology relation 
(Dressler 1980): early type galaxies are more frequently found in high density environments, i.e. galaxy clusters. At 
intermediate redshifts ($z \le 0.5$) the fraction of blue starforming galaxies in clusters is higher than that in local 
clusters (Butcher--Oemler effect; Butcher \& Oemler 1978, 1984). 
Together with the result that the fraction of S0 galaxies increases with decreasing redshift (Dressler et al. 1997; 
Fasano et al. 2000), this indicates that infalling spiral galaxies are transformed into S0 galaxies. Since morphology 
is closely related to the star formation history of a galaxy, it is not surprising that the average galaxy properties 
related to star formation also depend on local density (Hashimoto et al. 1998; Lewis et al. 2002; Gomez et al. 2003; 
Kauffmann et al. 2004; Balogh et al. 2004). The analysis of galaxies in ten distant clusters (Dressler et al. 1999; 
Poggianti et al. 1999) has shown that the galaxy populations of these clusters are characterized by the presence of a 
large number of post starburst galaxies. Poggianti et al. (1999) concluded that the most evident effect due to the 
cluster environment is the quenching of star formation rather than its enhancement. They found two different galaxy 
evolution timescales in clusters: (i) a rapid removal of the gas ($\sim$ 1 Gyr) and (ii) a slow transformation of 
morphology (several Gyr). A recent study of two massive, intermediate-redshift galaxy clusters (Moran et al. 2007) 
confirmed these findings: spiral galaxies within infalling groups already begin a slow process of conversion into S0s. 
Once they reach the cluster core, ram pressure stripping dominates where the intracluster medium is dense. Otherwise, 
the slow transformation process goes 
on. Whereas the slow transformation is most probably due to gravitational interactions (galaxy--cluster potential, Byrd \& 
Valtonen 1990, Valluri 1993; galaxy--galaxy; galaxy harassment, Moore et al. 1996, 1998), the rapid timescale is due to 
ram pressure stripping. 

The Virgo cluster of galaxies represents an ideal laboratory to study the effects of ram pressure stripping in detail. It is the only cluster in the northern hemisphere where one can observe the ISM distribution and kinematics of cluster galaxies at a kpc resolution (1~kpc $\sim 12''$)\footnote{I use a distance of 17~Mpc to the Virgo cluster}. The Virgo cluster is dynamically young and spiral-rich. Most of the cluster's spiral galaxies are H{\sc i} deficient, i.e. they have lost a significant amount of their ISM (Chamaraux et al. 1980, Giovanelli \& Haynes 1983). Imaging H{\sc i} observations have shown that these galaxies have truncated H{\sc i} disks ( Giovanelli \& Haynes 1983, Cayatte et al. 1990). Thus, the cluster environment changes the H{\sc i} content and morphology of Virgo cluster spiral galaxies. However, Virgo spiral galaxies are not CO-deficient (Kenney \& Young 1989).
In a new H{\sc i} imaging survey of Virgo galaxies (VIVA: VLA Imaging of Virgo galaxies in Atomic gas), Chung et al. (2007) found seven spiral galaxies with long H{\sc i} tails. These galaxies are found in intermediate- to low-density regions (0.6-1 Mpc in projection from M87). 
The tails are all pointing roughly away from M87, suggesting that these tails are due to ram pressure stripping. Therefore, ram pressure stripping already begins to affect spiral galaxies around the cluster Virial radius. The further evolution of a galaxy depends critically on its orbit (see, e.g., Vollmer et al. 2001), i.e. a highly eccentric orbit will lead the galaxy at a high velocity into the cluster core, where the intracluster medium is densest and ram pressure will be very strong.

In this article I will use the results from about 10 years of studying the effects on ram pressure stripping on individual Virgo cluster galaxies to (i) construct a first complete model-based time sequence and (ii) to verify if the model-deduced galaxy orbital segments are consistent with observations.
In Sect. 2 the method to derive  the ram pressure stripping parameters is described and  the uncertainties in the parameters are given,  the ram pressure stripping time sequence is established, followed by a description of the observational characteristics of the different stages of ram pressure stripping. In Sect.~\ref{sec:orbits} it is investigated if the model-derived parameters are consistent with eccentric galaxy orbits within the Virgo cluster. Sect.~\ref{sec:efficiency} is devoted to the ram pressure stripping efficiency, and the conclusions are given in Sect.~\ref{sec:conclusions}.

\section{The first complete model-based ram pressure stripping time sequence\label{sec:sequence}}

Within the last 6 years a sample of 6 Virgo cluster spiral galaxies was investigated in detail (Table~\ref{tab:galaxies} contains their B magnitudes, radial velocities, optical radii, and distances to M~87; for references see Table~\ref{tab:refs}) based on VLA H{\sc i} and IRAM 30m CO observations, and the dynamical models described in Vollmer et al. (2001). These galaxies are located at projected distances of 0.3--1.0~Mpc from the cluster center, M 87 (Fig.~\ref{fig:viva_rps}). Besides tell-tale signs in their H{\sc i} properties (see Sect. 2.4), all galaxies show asymmetric ridges of polarized radio continuum emission in the outer galactic disk (Vollmer et al. 2007), which  are tracers of interactions with their environment and can be used as diagnostic tools to determine the kind of interaction (gravitational or ram pressure) and the interaction parameters (Vollmer et al. 2006; Soida et al. 2006). 

\begin{table}
      \caption{Sample of Virgo spiral galaxies.}
         \label{tab:galaxies}
      \[
         \begin{tabular}{lllll}
           \hline
	   Name & B & $v_{\rm r}$ & $R_{25}$ & $d_{\rm M87}$ \\
	    & (mag) & (km\,s$^{-1}$) & ($'$) & (Mpc) \\
	   \hline
	   NGC\ 4330 & 14.0 & 1572 & 4.0 & 0.6 \\
	   NGC\ 4388 & 12.2 & 2515 & 5.5 & 0.4 \\
	   NGC\ 4438 & 12.0 & 98 & 8.9 & 0.3 \\
	   NGC\ 4501 & 10.6 & 2270 & 6.8 & 0.6 \\
	   NGC\ 4522 & 13.6 & 2324 & 3.5 & 1.0 \\
           NGC\ 4569 & 11.8 & 643& 10.2 & 0.5\\
	   \hline
        \end{tabular}
      \]
\end{table}

\subsection{Determining the stripping parameters \label{sec:method}}

In the following I describe how the stripping parameters are determined by comparing the dynamical simulations with observations.
For a direct comparison between model and observations of  a given galaxy the observed parameters are: its systemic radial velocity, projected distance from the cluster center, disk inclination angle, disk position angle, gas distribution, and velocity field. With the help of the dynamical model the maximum ram pressure $p_{\rm ram}^{\rm max}$,
the time to maximum ram pressure $t_{\rm rps}$, and the angle $i$ between the galactic disk and the ram pressure wind direction were determined in the following way:
\begin{enumerate}
\item
Maximum ram pressure and ram pressure wind inclination angle:\\
the galaxy's H{\sc i} deficiency ($\log(M_{\rm HI}^{\rm exp}/M_{\rm HI}^{\rm obs})$), where $M_{\rm HI}^{\rm exp}$ is the expected and $M_{\rm HI}^{\rm obs}$  the observed H{\sc i} mass, sets the quantity $p_{\rm ram}^{\rm max} \sin i$:
\begin{equation}
M_{\rm HI}^{\rm in}/M_{\rm HI}^{\rm f} \propto p_{\rm ram}^{\rm max}\,\sin^{2}(\frac{9}{10}(i +10^{\rm o}))
\end{equation}
(Eq.~21 of Vollmer et al. 2001), where $p_{\rm ram}^{\rm max}$ is the maximum ram pressure and $M_{\rm HI}^{\rm in/f}$ the initial and final H{\sc i} masses of the simulation, respectively. Typically, the H{\sc i} distribution of a stripped spiral galaxy shows a sharp outer edge on one side and more extended, extraplanar gas on  the opposite side. The determination of the inclination angle $i$ is mainly based on the morphology of the extraplanar gas. The derived maximum ram pressure is generally close to that derived from the Gunn \& Gott criterion (Gunn \& Gott 1972):
\begin{equation}
p_{\rm ram}^{\rm max} = \Sigma_{\rm ISM} v_{\rm rot}^{2} R^{-1}\ , 
\end{equation}
where $R$ is the galactocentric distance, $v_{\rm rot}$ the rotation velocity, and $\Sigma_{\rm ISM} \sim 10^{21}$~cm$^{-2}$ the ISM surface density at the sharp edge of the H{\sc i} distribution.
\item
Temporal ram pressure profile:\\
the profile is fixed by the maximum ram pressure due to the galaxy motion within the intracluster medium . Vollmer et al. (2001, 2008) calculated temporal ram pressure profiles for a galaxy on eccentric orbits within the gravitational potential of the Virgo cluster assuming an intracluster medium density derived from ROSAT X-ray observations (Schindler et al. 1999). Their correlation between the maximum ram pressure and the width of the temporal stripping profiles
is used for the dynamical models.
\item
Azimuthal viewing angle:\\
for the comparison with observations the simulated galaxy has to be projected onto the sky. The major axis position angle and inclination of the galaxy define a plane in three dimensional space. The simulated galaxy can then be rotated within this plane by an azimuthal viewing angle (see e.g. Vollmer et al. 2008).  The three-dimensional model wind direction, the line-of-sight velocity of the galaxy, and the projected ICM wind direction can  thus all be expressed as functions of the azimuthal viewing angle.
\item
Time to, or from ram pressure peak:\\
finally,  the model snapshot is chosen which reproduces best the gas morphology, velocity field, and observed line-of-sight velocity of the galaxy. This sets the time to or from ram pressure maximum, i.e. the stripping age. In practice, ram pressure simulations are done for
several different wind directions.
\end{enumerate}

\begin{table*}
      \caption{References.}
         \label{tab:refs}
      \[
         \begin{tabular}{lll}
           \hline
	    & observations & model \\
	   \hline
	   NGC~4501 & Vollmer et al. (2008) & Vollmer et al. (2008) \\
	   NGC~4330 & Chung et al. (2007) & Vollmer et al. (in prep.) \\
	    & Abramson et al. (in prep.) & \\
	   NGC~4438 & Vollmer et al. (2005) & Vollmer et al. (2005) \\
	   NGC~4522 & Kenney et al. (2004) & Vollmer et al. (2006) \\
	            & Vollmer et al. (2004b) & \\
	   NGC~4388 & Vollmer \& Huchtmeier (2003) & Vollmer \& Huchtmeier (2003) \\
	            & Oosterloo \& van Gorkom (2005) & \\
	   NGC~4569 & Vollmer et al. (2004a) & Vollmer et al. (2004a) \\
	   \hline
        \end{tabular}
      \]
\end{table*}

\subsection{Uncertainties of the model-derived interaction parameters \label{sec:uncertain}}

The uncertainties in the model-derived
interaction parameters can only be estimated, since the choice of the ``best-fit'' model is made by eye and only a few simulations are made. For models where the galaxy has already experienced the ram pressure peak (post-peak models), the maximum ram pressure has left its imprint on the gas distribution and velocity field. On the other hand, in the case of a pre-peak model, the peak ram pressure to be attained depends on assumptions on the galaxy orbit within the cluster. In the following I estimate the uncertainties of the interaction parameters based on the case of NGC~4522 (Vollmer et al. 2006) for post-peak, and on NGC~4501 (Vollmer et al. 2008) for pre-peak ram pressure stripping:
\begin{itemize}
\item
post-peak: the uncertainties in the peak ram pressure $p_{\rm ram}^{\rm max}$, time to peak ram pressure $t_{\rm rps}$, and inclination angle $i$ between the disk plane and the ram pressure wind direction are:
$\Delta p_{\rm ram}^{\rm max} \sim 20$~Myr, $\Delta t_{\rm rps} \sim 0.1\ t_{\rm rps}$, and $\Delta i \sim 10^{\circ}$.
\item
pre-peak: the uncertainty in $i$ is $\Delta i \sim 10^{\circ}$. Maximum ram pressure can only be determined if a galaxy orbit is assumed.
The uncertainties in  $p_{\rm ram}^{\rm max}$ and $t_{\rm rps}$ are then $\Delta p_{\rm ram}^{\rm max} \sim 20$~Myr and $\Delta t_{\rm rps} \sim 0.1\ t_{\rm rps}$.
\end{itemize}
I thus estimate the total uncertainties on the model-derived parameters to be $\Delta p_{\rm ram}^{\rm max}= 20$~Myr, $\Delta t_{\rm rps}= 0.1\ t_{\rm rps}$, and $\Delta i =10^{\circ}$. It should be noted that these uncertainties do not significantly affect the following results and conclusions.

\subsection{Observational gas characteristics}

Based on our H{\sc i} and CO observations the Virgo cluster spiral galaxies can be divided into the following 4 classes, based on their gas morphology. These classes can be arranged to form a time sequence for ram pressure stripping in the Virgo cluster (Fig.~\ref{fig:sequence}, as will be argued hereafter).

\begin{itemize}
\item
(i) gas disk truncated near the optical radius ($R_{25}$) and asymmetric outer gas disk with a tail structure (NGC~4501, NGC~4330);
\item
(ii) strongly truncated gas disk and extraplanar high surface density gas (NGC~4522, NGC~4438);
\item
(iii) strongly truncated gas disk and low surface brightness gas (NGC~4388);
\item
(iv) truncated gas disk and perturbed low surface brightness arms (NGC~4569).
\end{itemize}

\subsection{The ram pressure stripping time sequence \label{sec:time}}

\begin{figure*} 
\begin{center}
        \resizebox{\hsize}{!}{\includegraphics{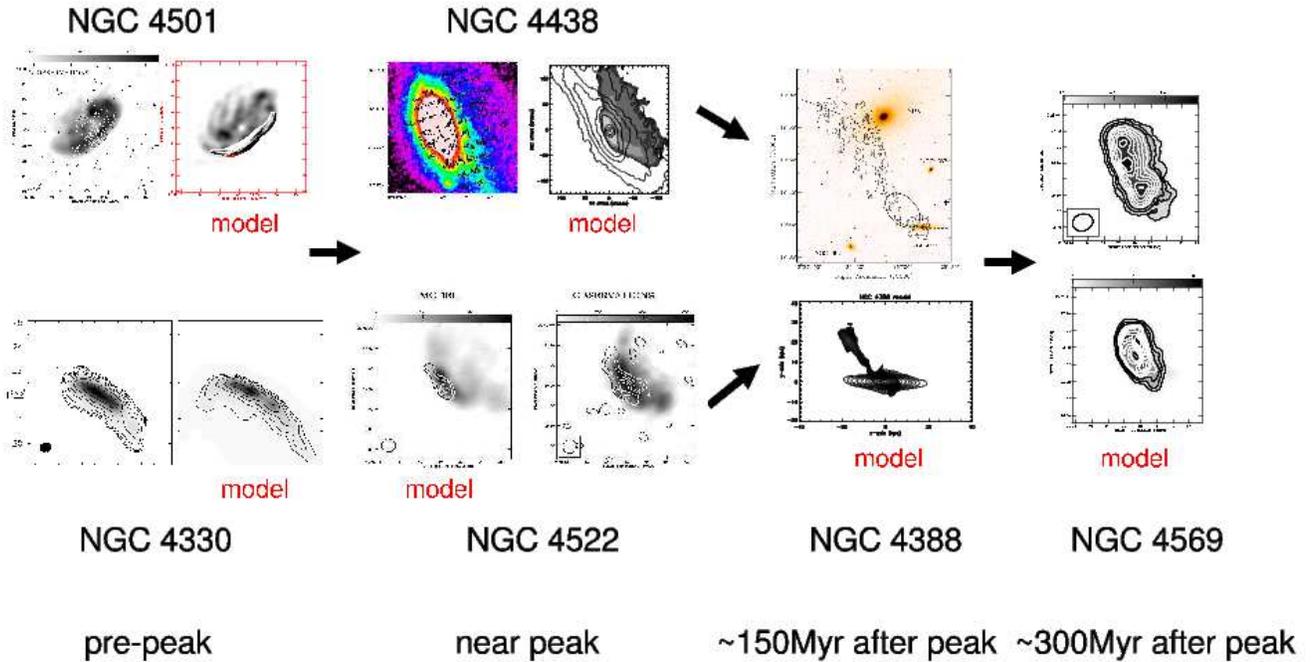}}
        \caption{Model-based complete ram pressure stripping time sequence for Virgo cluster spiral galaxies. NGC~4501 (greyscale: H{\sc i}, contour: polarized radio continuum emission; Vollmer et al. 2008) and NGC~4330 (greyscale and contours: H{\sc i}; Chung et al. 2007; Vollmer et al. in prep.) are approaching the cluster center and are thus in a stage of pre-peak ram pressure (class (i)). NGC~4522 (greyscale: H{\sc i}, contour: polarized radio continuum emission; Kenney et al. 2004; Vollmer et al. 2006) and NGC~4438 (observations: CO spectra on optical image; model: greyscale: gas surface density; contours: stellar distribution; Vollmer et al. 2005) are close to peak ram pressure (class(ii)). NGC~4388 (class(iii); observations: contour: H{\sc i} Oosterloo \& van Gorkom 2005; model: greyscale: gas surface density; contour: stellar distribution; Vollmer \& Huchtmeier 2003) and NGC~4569 (class (iv); greyscale and contours: H{\sc i}; Vollmer et al. 2004a) are leaving the cluster center. 	  
        } \label{fig:sequence}
\end{center}
\end{figure*}

The dynamical models tell us that class (i) galaxies are observed before peak ram pressure, class (ii) galaxies are close to peak ram pressure, and galaxies of classes (iii) and (iv) are observed after peak ram pressure (Fig.~\ref{fig:sequence}). 

Crowl \& Kenney (2008) independently confirmed the model-derived stripping ages (time to peak ram pressure)  by analyzing the stellar populations of the stripped, gas-free outer parts of the disks of NGC~4388, NGC~4522, and NGC~4569.

A more detailed description of the different stripping stages is:
\begin{itemize}
\item
pre-peak (class (i)):\\
The galaxies  (NGC~4501 and NGC~4330) are at the beginning of a ram pressure stripping event (Vollmer et al. 2008; Vollmer et al. in prep.), i.e. they are approaching the cluster center. The outer gas disk has already been removed and the gas of the inner disk is just beginning to be affected by ram pressure. 
\item
near peak (class (ii)):\\
NGC~4438 is at present undergoing strong ram pressure stripping. Despite the strong tidal perturbation, ram pressure is the dominant effect on the observed gas distribution and kinematics (Vollmer et al. 2005). NGC~4522 is also close to peak ram pressure. The gas disks of both galaxies are strongly truncated and the extraplanar gas has a high column density.
\item
post-peak (class (iii) and (iv)):\\
after peak ram pressure, i.e when the galaxy leaves the cluster core again, the removed ISM is accelerated and expands. The extraplanar gas tail becomes larger and its gas surface density decreases. This is observed in NGC~4388 (Oosterloo \& van Gorkom 2005) where the dynamical model yields a time to peak ram pressure of $\sim 100-200$~Myr.

About $200$~Myr later the gas tail is still expanding, its surface density is decreasing, and the gas is evaporated by the hot intracluster medium.  The gas tail is thus no longer detectable in H{\sc i}. Ram pressure is decreasing and the pushed gas, which has not been accelerated
to the escape velocity, falls back onto the galaxy and begins to resettle there. Thus, the only traces of the past interaction are kinematically
perturbed gas arms located at the outer edge of the gas disk due to resettling gas, as observed in NGC~4569.
\end{itemize}

\subsection{How to recognize ram pressure stripped galaxies \label{sec:smoke}} 

From the work on the 6 galaxies (Table~\ref{tab:refs}) I have identified the following observational characteristics (``smoking guns'') for the different stages of ram pressure stripping:
\begin{itemize}
\item
pre-peak ram pressure stripping:\\
in the case of the detection of extraplanar gas, increasing, moderately strong ram pressure ($p_{\rm ram} \sim 5000$~cm$^{-3}$km\,s$^{-1}$) leads to extraplanar gas of moderate mean surface density (several $10^{19}$~cm$^{-2}$) close to the galactic disk with a continuous velocity field from the disk to the extraplanar region (NGC~4501, NGC~4330).
\item
near peak ram pressure stripping:\\
ongoing strong ram pressure stripping ($p_{\rm ram} \ga 20000$~cm$^{-3}$km\,s$^{-1}$) gives rise to high mean surface density (several $10^{20}$~cm$^{-2}$) extraplanar gas close to the galactic disk with a continuous velocity field from the disk to the extraplanar region (NGC~4522, NGC~4438).
\item
post-peak ram pressure stripping:\\
for intermediate stripping ages ($100$--$200$~Myr after peak) gas can be found at large distances from the galactic disk. The mean surface density of this extraplanar gas is low  (a few $10^{19}$~cm$^{-2}$; NGC~4388).
The velocity field is continuous from the disk to the extraplanar region.

For higher stripping ages ($>200$~Myr) perturbed gas arms are visible at the outer edge of the gas disk showing a discontinuous velocity field from the disk to the extraplanar region (NGC~4569).

At the very end of this sequence ($> 300$~Myr) stripped spiral galaxies show a truncated symmetric and unperturbed gas disk.

\end{itemize}

Galaxies with stripping ages $<200$~Myr show ridges of polarized radio continuum emission at the outer edge of the disk opposite to the extraplanar gas. These ridges are due to ram pressure compression of the large-scale magnetic field. Galaxies which are observed more than 200~Myr after peak ram pressure can also show enhanced polarized radio continuum emission due to shear motions of re-accreting or resettling gas
(Otmianowska-Mazur \& Vollmer 2003).

The time window during which one can still identify perturbations due to ram pressure stripping is $\sim 500$~Myr. The orbital period of the orbits shown in Fig.~\ref{fig:orbits} is several Gyr. This relatively short time window for the detection of ram pressure induced perturbations explains the rareness of H{\sc i} tails in the VIVA sample (7 out of 50 galaxies; Chung et al. 2007).

\section{Orbits of ram pressure stripped galaxies\label{sec:orbits}}

The ram pressure stripping time sequence presented in Sect.~\ref{sec:time} does not depend on any assumption on galaxy orbits. It only relies on the model-derived time to peak ram pressure (pre-peak, near-peak, or post-peak). In the following, I investigate if the model-derived parameters are consistent with realistic galaxy orbits within the Virgo cluster, given the positions, velocities and directions of the galaxies within the cluster.

\begin{figure}
\begin{center}
        \resizebox{\hsize}{!}{\includegraphics{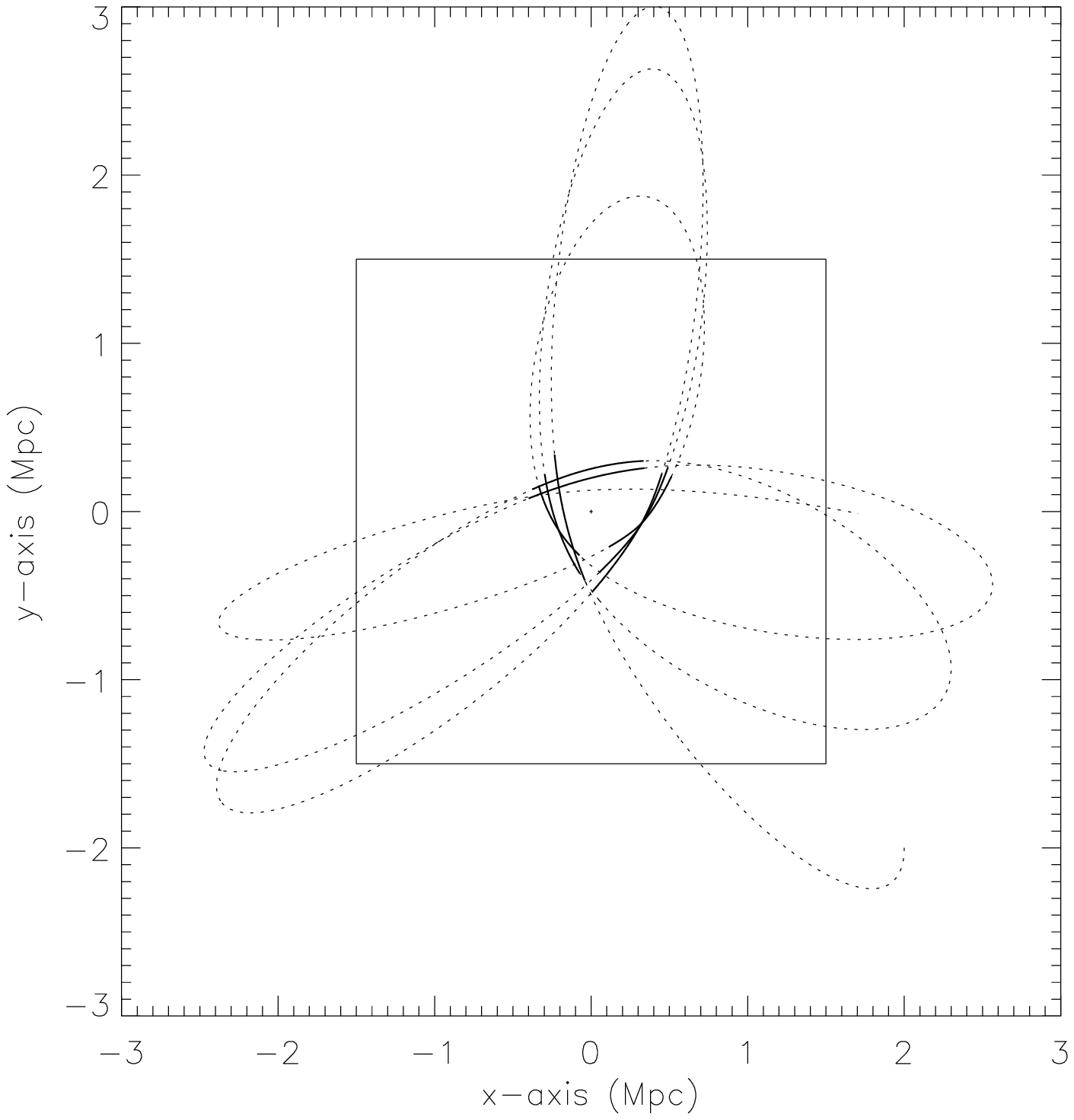}}
        \caption{Galaxy orbits in the Virgo cluster with peri- to apocenter ratios of about 1:10, giving rise to ram pressure maxima between $10\,p_0$ and $50\,p_0$, where the normalisation is $p_0=\rho_0 v_0^2$ with $\rho_{0}$=10$^{-4}$ cm$^{-3}$ and $v_{0}$=1000~km\,s$^{-1}$. The inner box delimitates the extent of Fig.~\ref{fig:viva_rps}.
        } \label{fig:orbits}
\end{center}
\end{figure}

The dynamical model yields the 3D velocity vector of the galaxy, the peak ram pressure, and the time to peak ram pressure. I assume linear orbits over an interval of $\pm 200$~Myr around peak ram pressure. Fig.~\ref{fig:orbits} shows typical orbits in the gravitational potential of the Virgo cluster (Schindler et al. 1999), confirming this assumption as a zero-order approximation. To calculate distances, I further estimate a mean galaxy velocity $v$ on this linear trajectory with respect to the cluster mean, which is based on the peak ram pressure and the time to peak ram pressure from the galaxy orbit simulations of Vollmer et al. (2001) and on the simulation shown in 
Fig.~\ref{fig:orbits}. The uncertainty in this mean velocity is about $\Delta v \sim 0.2 v$.

In the case of a smooth, static, and spherical intracluster medium, peak ram pressure occurs at the galaxy's closest approach to the cluster center, i.e. when the galaxy's velocity vector is perpendicular to its distance vector. This condition allows the calculation of the galaxy's present line-of-sight distance and its 3D position during peak ram pressure (Table~\ref{tab:numbers})\footnote{Negative numbers indicate directions to the left, bottom, and in front of M~87.}:
let the galaxy's 3D velocity vector be $\vec{v}$ and the present position $\vec{x}=(x,y,z)$, where the line-of-sight component $z$ is unknown. Within a time $\Delta t$ the galaxy moves a distance of $\Delta \vec{s} = \vec{v} \Delta t$. The galaxy's position at peak ram pressure is then $\vec{x_0}=(x_0,y_0,z_0)=\vec{x} + \Delta \vec{s}$.
During peak ram pressure the velocity and the distance vectors are perpendicular, i.e. $\vec{v} \cdot \vec{x_0} = 0$. In this way $z_0$ and subsequently $z$ and the minimum distance to the cluster center at peak ram pressure can be calculated. Since I estimate the uncertainty in the direction of the model galaxy's 3D velocity vector to be $\sim 10^{\circ}$, I searched for the direction leading to the minimum impact parameter (closest distance to the cluster center) within a $10^{\circ}$-cone around the model velocity vector.

\begin{figure}
\begin{center}
        \resizebox{\hsize}{!}{\includegraphics{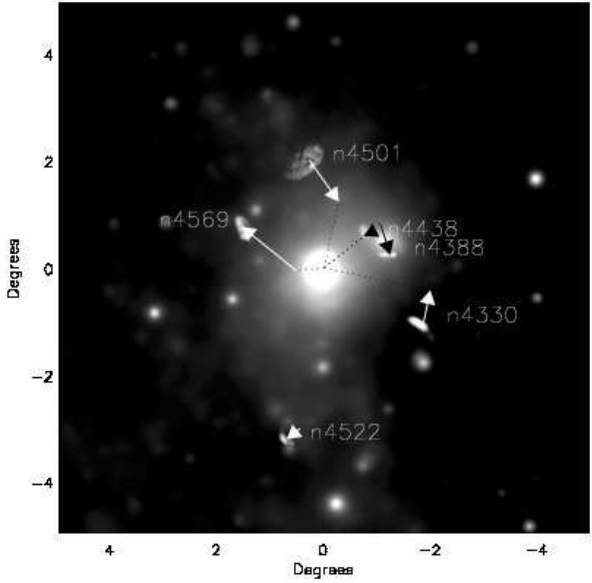}}
        \caption{Orbital segments from the location of peak ram pressure to the present location of the 6 Virgo cluster spiral galaxies. A spherical, smooth, and static intracluster medium is assumed. The arrows indicate the galaxies' motions projected on the sky. Linear orbital segments are assumed. The dotted lines connect the location of peak ram pressure and the cluster center (M~87). The 3D orbital segments are shown in Fig.~\ref{fig:orbits}.
        } \label{fig:viva_rps}
\end{center}
\end{figure}

\begin{figure}
\begin{center}
        \resizebox{8cm}{!}{\includegraphics{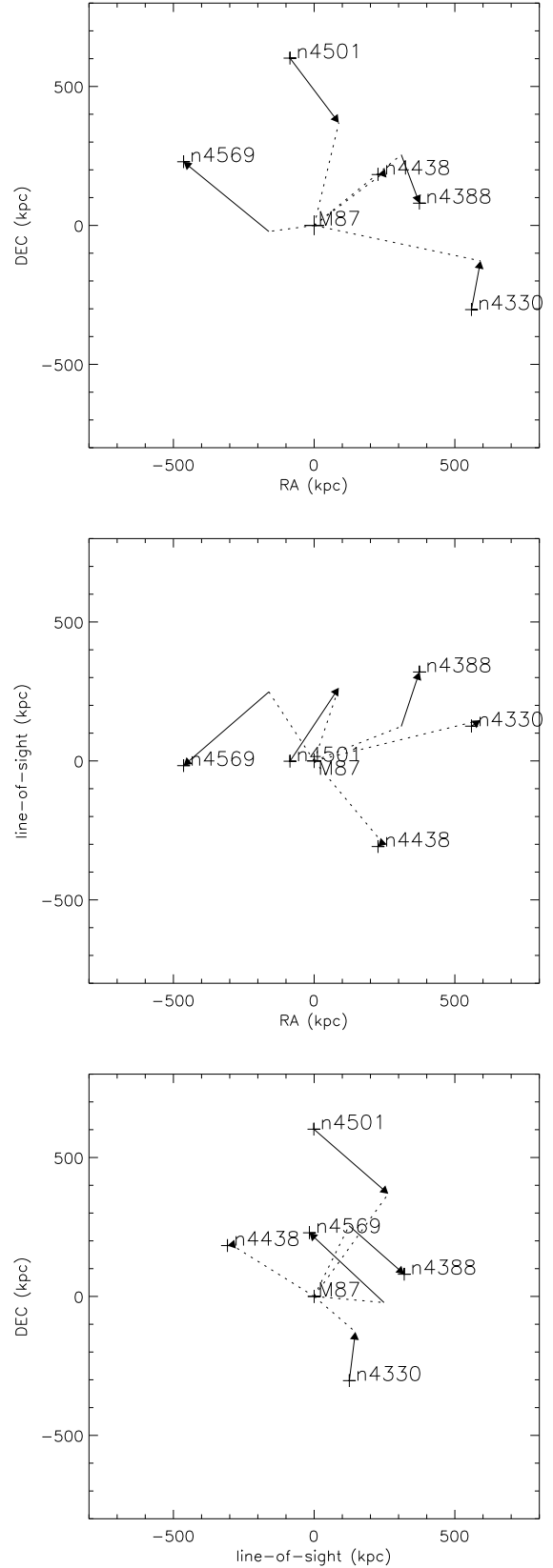}}
        \caption{Three dimensional linear orbital segments of the 6 Virgo cluster spiral galaxies. The dotted lines connect the location of peak ram pressure and the cluster center (M~87). A spherical, smooth, and static intracluster medium is assumed. The orbital segment of NGC~4438 is questioned in Appendix A.
        } \label{fig:holistic2}
\end{center}
\end{figure}

Changing the galaxy's mean velocity by 20\% leads to a 20\% variation in the galaxy's 3D position. A change in the time to peak ram pressure has the same effect. The largest variation of the resulting orbital parameters is observed when the galaxy's 3D velocity vector is modified. This variation depends strongly on the initial, model-derived velocity vector. Whereas the variation in the galaxy's impact parameter $D_{\rm min}$ due to a change of $10^{\circ}$ in the velocity vector is modest ($\la 20$\%) for NGC~4501, NGC~4438, NGC~4388, and NGC~4569, it is a factor of 2 for NGC~4330. The resulting galaxy distances from the cluster center along the line-of-sight vary by $\sim 100$~kpc for these four galaxies, and by $1$~Mpc for NGC~4330. Altogether, I estimate the uncertainty of the galaxy's smallest distance to M~87 to be $\sim 30$\%.
The linear orbital segments determined in this way together with the minimum distances to the cluster center are presented in Fig.~\ref{fig:viva_rps} and \ref{fig:holistic2}.

The galaxies have random velocity vectors within the cluster. The range of the line-of-sight positions is somewhat smaller, but comparable to the ranges along the right ascension and declination axes. NGC~4569 has the smallest ($210$~kpc), NGC~4330 the largest ($620$~kpc) impact parameter. 
\begin{table*}
      \caption{Orbital parameters with respect to M~87. Negative numbers indicate directions to the left, bottom, and in front of M~87. Row (1)--(3): present galaxy position; row (4)--(6): 3D unit vector of the galaxy's motion within the cluster from the dynamical models; row (7)--(9): 3D unit velocity vector leading to the minimum impact parameter $D_{\rm min}$; row (10): total galaxy velocity; row (11): normalized ram pressure $p_{\rm norm}=(\rho_{\rm ICM} v_{\rm galaxy}^{2})/(\rho_{0} v_{0}^{2})$, where $\rho_{0}$=10$^{-4}$ cm$^{-3}$ and $v_{0}$=1000 km\,s$^{-1}$; row (12): time to peak ram pressure (negative values are before peak ram pressure); row (13)--(15): galaxy position at peak ram pressure assuming a spherical, smooth, and static intracluster medium; row (16): impact parameter or distance at peak ram pressure. In Appendix A it is argued that NGC~4438 does most probably not follow the given orbit but is rather located $\sim 1$~Mpc behind M~87. Since NGC~4522 is being stripped by a moving intracluster medium, its  impact parameter cannot be determined. The uncertainties in the parameters are discussed in Sect.~\ref{sec:uncertain}.}
         \label{tab:numbers}
      \[
         \begin{array}{lrrrrrr}
           \hline
	    & {\rm NGC}~4501 & {\rm NGC~4330} & {\rm NGC~4438} & {\rm NGC~4388} & {\rm NGC~4569} & {\rm NGC~4522} \\
	   \hline
	   x\ {\rm (kpc)} & -86 & 559 & 227 & 374 & -464 & -210 \\
	   y\ {\rm (kpc)} & 602 & -303 & 183 & 80 & 229 & -955 \\
	   z\ {\rm (kpc)} & -1 & 125 & (-308) & 320 & -17 & (500) \\
	   n_{\rm x}^{\rm mod} & 0.48 & 0.05 & -0.81 & 0.09 & -0.64 & -0.73 \\
	   n_{\rm y}^{\rm mod} & -0.44 & 0.99 & -0.20 & -0.75 & 0.53 & -0.60 \\
	   n_{\rm z}^{\rm mod} & 0.76 & 0.08 & -0.55 & 0.66 & -0.56 & 0.33 \\
	   n_{\rm x} & 0.44 & 0.18 & (-0.72) & 0.24 & -0.75 & - \\
	   n_{\rm y} & -0.59 & 0.98 & (-0.13) & -0.65 & 0.52 & - \\
	   n_{\rm z} & 0.67 & 0.12 & (-0.68) & 0.73 & -0.41 & - \\	   
	   v\ ({\rm km\,s}^{-1}) & 1500 & 1700 & (2000) & 1700 & 1500 & 2000 \\
	   p_{\rm norm} & 20 & 20 & 50 & 50 & 40 & 50 \\
	   \Delta\ t\ {\rm (Myr)} & -250 & -100 & 10 & 150 & 300 & 50 \\
	   x_0\ {\rm (kpc)} & 88 & 592 & (242) & 309 & -108 & - \\
	   y_0\ {\rm (kpc)} & 368 & -128 & (186) & 255 & -15 & - \\
	   z_0\ {\rm (kpc)} & 266 & 146 & (-294) & 127 & 178 & - \\
	   D_{\rm min}\ {\rm (kpc)} & 460 & 620 & (420) & 420 & 210 & - \\
	   \hline
        \end{array}
      \]
\end{table*}

\section{The efficiency of ram pressure stripping\label{sec:efficiency}}

It is now possible to check the consistency between the orbital segments deduced in Sect.~\ref{sec:orbits} and the maximum ram pressure determined from the dynamical models (Table~\ref{tab:refs} and \ref{tab:numbers}).
The link between these two quantities is the density distribution of the intracluster medium, again assuming that it is smooth, static, and spherical. Schindler et al. (1999) determined the intracluster medium density distribution in the Virgo cluster from ROSAT X-ray observations,
assuming hydrostatic equilibrium:
\begin{equation}
\rho=\rho_{\rm C}\big(  1+ \frac{r^{2}}{r_{\rm C}^{2}}\big) ^{-\frac{3}{2}\beta}\label{eq:betamodel}
\end{equation} 
where $r_{\rm C}$ is the core radius and $\rho_{\rm C}$ is the central density. 
I assume $\rho_{\rm C}=4.2 \times 10^{-2}$~cm$^{-3}$, $r_{\rm C}=13.4$~Mpc, and $\beta=\frac{7}{15}$ (Schindler et al. 1999).
For the galaxy velocity at peak ram pressure I assume $v=2000$~km\,s$^{-1}$ based on the galaxy orbit simulations of Vollmer et al. (2001) and the simulation shown in Fig.~\ref{fig:orbits}.
I define the ram pressure efficiency as a constant $\xi$ with $p_{\rm ram}=\xi \rho_{\rm ICM} v^2$, where $\rho_{\rm ICM}$ is the intracluster medium density and $v$ the galaxy velocity with respect to the intracluster medium. In previous works $\xi=1$ has been assumed.

\begin{figure}
\begin{center}
        \resizebox{\hsize}{!}{\includegraphics{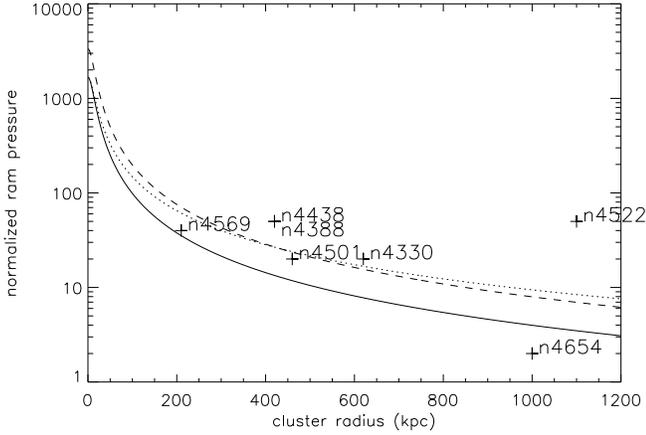}}
        \caption{Normalized ram pressure ($p_{\rm norm}=(\rho_{\rm ICM} v_{\rm galaxy}^{2})/(\rho_{0} v_{0}^{2})$, where $\rho_{0}$=10$^{-4}$ cm$^{-3}$ and $v_{0}$=1000 km\,s$^{-1}$) as a function of cluster radius. The solid line traces the normalized ram pressure assuming the intracluster medium density distribution $\rho_{\rm ICM}$ (Eq.~\ref{eq:betamodel}) of Schindler et al. (1999) and $v=2000$~km\,s$^{-1}$. The normalized ram pressure with $2\,\rho_{\rm ICM}$ (dashed line) and a different $\beta$ (dotted line) are also shown. The distance of NGC~4654 is a lower limit.
        } \label{fig:holistic1}
\end{center}
\end{figure}

The ram pressure $p_{\rm ram} = \rho_{\rm ICM}  v^2$ can be seen together with the values derived for our 6 sample galaxies in Fig.~\ref{fig:holistic1}. Except for NGC~4569 the assumed intracluster medium density distribution leads to a too small maximum ram pressure compared to the values derived from the dynamical models. 
Enhancing the intracluster medium density distribution by a factor of 2 or decreasing $\beta=\frac{2}{5}$ leads to more consistent results. 
Thus, I conclude that either (i) the stripping efficiency and/or (ii) the intracluster medium density has been underestimated.

Compared to NGC~4330, NGC~4501, and NGC~4569, the two spirals which are located close to M~86, NGC~4388 and NGC~4438, require a still 2 times higher peak ram pressure than expected from a smooth and static intracluster medium assuming a higher stripping efficiency and/or a higher intracluster medium density (see Appendix A).

For NGC 4388, increasing the time from peak ram pressure from 150 Myr to $200$~Myr as advocated by Crowl \& Kenney (2008)
based on an optical spectrum and UV flux of the gas-free outer disk, would increase its impact parameter from 420~kpc to 460~kpc. Decreasing the time from peak ram pressure to $100$~Myr (Vollmer \& Huchtmeier 2003), would decrease the impact parameter by less than 10\,\%. Allowing for a larger uncertainty in the galaxy's velocity vector yields an impact parameter of 310~kpc for the direction $\vec{n}=(0.54,-0.48,0.70)$. The angle between this direction and that 
of Table~\ref{tab:numbers} is $18^{\circ}$ in the plane of the sky.
I conclude that for NGC~4388 either the intracluster medium is locally a factor of 2 denser than predicted by a spherical, smooth, and static intracluster medium distribution, or that the galaxy's unit velocity vector is  $\sim 20^{\circ}$ different from the values given in Table~\ref{tab:numbers}.

The cause for the unusual extent of NGC~4388's H{\sc i} tail and the complicated case of NGC~4438 are discussed in Appendix A. There I advocate a scenario where NGC~4438 has been stripped by the hot X-ray gas halo associated with M~86, has recently left  M~86's Mach cone,
and is now being stripped by the intracluster medium of the Virgo cluster. This scenario implies that NGC~4438's orbit is different from that given by Table~\ref{tab:numbers}. Its trajectory is close to M~86,
i.e. $\sim 1$~Mpc behind M~87. In this case, NGC~4438 cannot be used to test the consistency between galaxy orbits, dynamical models, and observations of the galaxy's ISM.

For comparison I have also added NGC~4522 and NGC~4654 to Fig.~5. As demonstrated in Kenney et al. (2004), the peak ram pressure for NGC~4522 is far too high to be explained in our smooth and static intracluster medium scenario. Indeed, Kenney et al. (2004), Vollmer et al. (2004),
and Vollmer et al. (2006) claim that NGC~4522 is encountering a moving intracluster medium which is presumably associated to M~49.  Therefore, its line-of-sight distance cannot be determined. Since the galaxy subcluster associated with M~49 is falling into the Virgo cluster from behind (Irwin \& Sarazin 1996; Biller et al. 2004) and the intracluster medium of M~49 is interacting with the intracluster medium of M87 (Shibata et al. 2001), I have assumed a line-of-sight distance of $z=500$~Mpc. Based on the dynamical model, NGC~4654 experiences a very low ongoing ram pressure together with a gravitational interaction which occurred in the recent past (Vollmer 2003). Again, its line-of-sight distance cannot be determined and the plotted value is a lower limit. The low ram pressure indicated that NGC~4654 might be located as far as $\sim 2$~Mpc behind the cluster center (M~87).

I conclude that the efficiency of ram pressure stripping derived from a sample of 4 Virgo spiral galaxies (NGC~4501, NGC~4330, NGC~4388, and NGC~4569) is about unity. The ram pressure stripping efficiencies of NGC~4522 and NGC~4654 are consistent with being close to unity.

\section{Conclusions\label{sec:conclusions}}

A first complete ram pressure stripping time sequence could be established by combining the results of detailed comparisons between dynamical models and observations of the interstellar medium in ram pressure-stripped galaxies in the Virgo cluster. It is possible to observe ram pressure induced perturbations $\sim 300$~Myr around
the galaxy's closest approach to the cluster center, i.e. when peak ram pressure occurs if a spherical, smooth, and static intracluster medium distribution is assumed.  The relative brevity of this period compared to the galaxy's orbital timescale (several Gyr) explains the rareness ($\sim$14\%) of observed ram pressure induced perturbations in Virgo spiral galaxies.

Observationally the different stages of ram pressure stripping can be recognized in the following way:
\begin{itemize}
\item
increasing, moderately strong ram pressure ($>50$~Myr before peak):\\
moderately truncated H{\sc i} disk, extraplanar gas of moderate surface density, continuous velocity field between the disk and the extraplanar region, ridge of polarized radio continuum emission at the outer gas disk opposite to the extraplanar region.
\item
ongoing strong ram pressure (near peak):\\
strongly truncated H{\sc i} disk, extraplanar gas of high surface density, continuous velocity field between the disk and the extraplanar region, ridge of polarized radio continuum emission at the outer gas disk opposite to the extraplanar region.
\item
decreasing ram pressure ($<200$~Myr after peak):\\
strongly truncated H{\sc i} disk, extended extraplanar gas of low surface density, continuous velocity field between the disk and the extraplanar region, ridge of polarized radio continuum emission at the outer gas disk opposite to the extraplanar region.
\item
decreasing ram pressure ($>200$~Myr after peak):\\
strongly truncated H{\sc i} disk, perturbed outer gas arms,
discontinuous velocity field between the disk and the extraplanar region,
ridge of polarized radio continuum emission at the outer gas disk opposite to the extraplanar region, possible ridge of polarized radio continuum emission at the outer gas disk due to shear motions from the resettling gas.
\end{itemize}

The dynamical models yield the 3D velocity vector of the galaxies, the peak ram pressures, and the times to peak ram pressure. In the case of a smooth, static, and spherical intracluster medium, peak ram pressure occurs during the galaxy's closest approach to the cluster center, i.e. when the galaxy's velocity vector is perpendicular to its distance vector. Under these conditions the galaxy's present line-of-sight distance and its 3D position during peak ram pressure can be calculated. The linear orbital segments derived in this way together with the intracluster medium density distribution from Schindler et al. (1999)
are consistent within a factor of 2 with the dynamical simulations for NGC~4501, NGC~4330, and NGC~4569. To resolve the discrepancy either a 2 times higher intracluster medium density than derived from X-ray observations by Schindler et al. (1999) and/or a 2 times higher stripping efficiency than assumed by the dynamical models, is needed. The impact parameters of the galaxy orbits vary between $200$~kpc and $600$~kpc. Compared to NGC~4501, NGC~4330, and NGC~4569, NGC~4388 requires a still 2 times higher local intracluster medium density or a slightly different direction of the galaxy's motion as derived from the dynamical model. 

It is suggested that NGC~4438 has moved within M~86's stripped hot X-ray gas during at least $\sim 100$~Myr, flew through the supposedly ram pressure stripped X-ray tail of M~86, recently left M~86's Mach cone, and is now encountering the Virgo intracluster medium. In this scenario the trajectory of NGC~4438 is located $\sim 200$~kpc behind M~86 and the interaction with NGC~4435 is responsible for NGC~4438's prominent tidal arms.

\begin{acknowledgements}
  I would like to thank Aeree Chung for providing me the background image of Fig.~\ref{fig:viva_rps}, Jacqueline van Gorkom and Jeffrey Kenney for fruitful discussions, and the anonymous referee and Wim van Driel who helped to significantly improve this article.
\end{acknowledgements}

\begin{appendix}

\section{The region around M~86}

As shown in Sect.~\ref{sec:efficiency}, the proposed impact parameters of the two galaxies closest to M~86, NGC~4438 and NGC~4388, are not consistent with the peak ram pressure deduced from the dynamical models. These discrepancies can be due to the parameters deduced from the dynamical models. For NGC~4388 the local intracluster medium density can well be enhanced by a factor of 2. I will explore the inconsistencies further in this Appendix.

\subsection{The tidal interaction of NGC~4438}

The case of NGC~4438 is more complicated than that of NGC~4388.  Dynamical models suggest that NGC~4438 had a gravitational interaction with NGC~4435 (Combes et al. 1988, Vollmer et al. 2005). Recent deep H$\alpha$ observations of the M~86 region (Kenney et al. 2008) revealed a highly complex and disturbed interstellar/intracluster medium. NGC~4438 is connected to M~86 by several faint H$\alpha$ filaments, which  might suggest a tidal interaction between the two galaxies, as advocated by Kenney et al. (2008). The timescale and the relative galaxy velocities of the two scenarios (encounter with M~86 or NGC~4435) are about the same. The higher mass of M~86 compared to that of NGC~4435 allows a larger impact parameter for an NGC~4438 -- M~86 encounter. Since M~86 has its own hot X-ray gas halo, the need for strong ongoing ram pressure that is presently affecting the ISM of the inner galactic disk, as shown in  Vollmer et al. (2005, 2008), implies that ram pressure stripping due to M~86's hot X-ray gas certainly did not remove, and might not even have significantly affected the ISM of NGC~4438. The recent discovery of a $\sim 190$~Myr-old starburst in NGC~4435 by Panuzzo et al. (2007) favors the interaction scenario between NGC~4438 and NGC~4435.

\subsection{The trajectory of NGC~4438}

In the following I assume that the faint H$\alpha$ filaments connecting NGC~4438 to M~86 indicate NGC~4438's direction of motion with respect to M~86. The model velocity vector of NGC~4438 (in km\,s$^{-1}$) is $\vec{v}_{\rm N4438}=(-1440,-260,-1360)\footnote{negative numbers indicate directions to the left, bottom, and in front of M~87}$ (Table~\ref{tab:numbers}). According to Finoguenov et al. (2004) and Randall et al. (2008), M~86 has a significant velocity component towards the south. 

The 3D velocity vector of M~86 is $\vec{v}_{\rm M86}=(v_x,v_y,-1500)$~km\,s$^{-1}$. NGC~4438's velocity vector within the infalling subcluster of M~86 is thus $\Delta \vec{v} = \vec{v}_{\rm N4438} - \vec{v}_{\rm M86} = (-1440-v_x,0.2(-1440-v_x),140)$~km\,$^{-1}$, where the $y$-component is calculated to fit the direction of the faint H$\alpha$ filaments connecting NGC~4438 to M~86. In this way I obtain $v_y=-260-0.2(1440+v_x)$~km\,$^{-1}$. Following the Randall et al. model leads to $v_x=v_y=-457$~km\,s$^{-1}$. M~86's total velocity with respect to M~87 is thus $1630$~km\,s$^{-1}$. NGC~4438's relative velocity vector with respect to M~86 is $\Delta \vec{v} = (-983,197,140)$~km\,s$^{-1}$. Consequently, NGC~4438 has evolved until recently on a west--east trajectory within the hot X-ray gas of M~86 at a speed of $\sim 1000$~km\,s$^{-1}$.

Following the Mach cone geometry of Randall et al. (2008) and assuming that NGC~4438 is just leaving the Mach cone yields a line-of-sight distance between NGC~4438 and M~86 of $\sim 200$~kpc. This is smaller than the physical length of M~86's X-ray tail (380~kpc; Randall et al. 2008).

\subsection{Ram pressure stripping of NGC~4438}

To estimate the ram pressure exerted on NGC~4438 by M~86's hot X-ray gas, its gas density needs to be known. Randall et al. (2008) estimate the density in M86's bright X-ray plume to be $n \sim 10^{-3}$~cm$^{-3}$. The gas density at larger distances from M~86 might thus be a few $10^{-4}$~cm$^{-3}$. On the other hand, the model deduced ram pressure for NGC~4438 (Vollmer et al. 2005) yields an intracluster medium density of $\sim 10^{-3}$~cm$^{-3}$ if a galaxy velocity of $2000$~km\,s$^{-1}$ is assumed.

In the proposed scenario NGC~4438 has been moving during the last $100$~Myr through the hot X-ray gas of  M~86 ($n \sim 10^{-4}$~cm$^{-3}$), has just recently (a few $10$~Myr ago) left M86's Mach cone (see Fig.~13 of Randall et al. 2008 for an illustration), and is presently encountering the Virgo cluster intracluster medium ($n \sim 10^{-3}$~cm$^{-3}$). Consequently, NGC~4438 is located at the same line-of-sight distance as M~86, i.e. $1.0 \pm 0.8$~Mpc behind M~87 (Mei et al. 2007). At the position of NGC~4438 B\"{o}hringer et al. (1994) have observed a filament of intracluster gas. Finoguenov et al. (2004) estimate the gas density in the filament to be $\sim 10^{-3}$~cm$^{-3}$. 

I conclude that NGC~4438 is not stripped by the spherical component of the Virgo intracluster medium. The stripping rather occurs at a physical distance from M87 between $500$~kpc, assuming a line-of-sight distance between M~86 and M~87 of $200$~kpc (Randall et al. 2008), and $\sim 1$~Mpc, adopting the M~86--M~87 distance from Mei et al. (2007). Together with NGC~4522, NGC~4438 is the second galaxy which is undergoing strong ram pressure stripping at a location in the Virgo cluster where this would  not be expected from a classical model assuming a spherical, smooth, and static intracluster medium distribution.

\subsection{The ram pressure stripped gas tail of NGC~4388}

Given that the 100~kpc long H{\sc i} tail of NGC~4388 is exceptional\footnote{There are two other prominent gas tails detected in NGC~4532/DDO 137 (Koopmann et al. 2008) and NGC~4254 (Haynes et al. 2007). Both are most probably of tidal nature.} for a galaxy at a projected distance from M~87 less than 0.5~Mpc, I also consider a possible ram pressure stripping event with M~86's intracluster medium within the Mach cone $\sim 200$~Myr ago as proposed by Oosterloo \& van Gorkom (2005). This would explain why such long H{\sc i} tails are not observed in the other galaxies of the present sample. An argument against this hypothesis is the extreme difference between the radial velocities of NGC~4388 and M~86 ($2700$~km\,s$^{-1}$) compared to the relatively low tangential velocity ($80$~kpc/$150$~Myr=$500$~km\,s$^{-1}$) which implies a ratio between the line-of-sight and the projected tail length of $\sim 5$ and thus a physical length of $\sim 400$~kpc. Another possibility might be that the gravitational potential of M~86 enhances locally the intracluster medium density near M~87 and the embedded gas of NGC~4388's stripped tail, leading to an enhanced recombination of the ionized gas in the tail which makes it observable in H{\sc i}.

\end{appendix}

\end{document}